\documentclass[a4paper]{jpconf}
\usepackage{graphicx}
\begin{document}
\title{\emph{Ab initio} investigations of $A{=}8$ nuclei: $\alpha{-}\alpha$ scattering, deformation in $^8$He, radiative capture of protons on $^7$Be and $^7$Li and the X17 boson}

\author{P. Navr\'atil}
\address{TRIUMF, Vancouver, British Columbia V6T 2A3, Canada}

\author{K. Kravvaris}
\address{Lawrence Livermore National Laboratory, P.O. Box 808, L-414, Livermore, CA 94551, USA}

\author{P. Gysbers}
\address{TRIUMF, Vancouver, British Columbia V6T 2A3, Canada}
\address{Department of Physics and Astronomy, University of British Columbia, Vancouver, British Columbia, V6T 1Z1, Canada}

\author{C.~Hebborn}
\address{Facility for Rare Isotope Beams, East Lansing, MI 48824, USA}
\address{Lawrence Livermore National Laboratory, P.O. Box 808, L-414, Livermore, CA 94551, USA}

\author{G.~Hupin}
\address{Université Paris-Saclay, CNRS/IN2P3, IJCLab, 91405 Orsay, France}

\author{S. Quaglioni}
\address{Lawrence Livermore National Laboratory, P.O. Box 808, L-414, Livermore, CA 94551, USA}

\ead{navratil@triumf.ca}

\begin{abstract}
We apply the No-Core Shell Model with Continuum (NCSMC) that is capable of describing both bound and unbound states in light nuclei in a unified way with chiral two- and three-nucleon interactions as the only input. The NCSMC can predict structure and dynamics of light nuclei and, by comparing to available experimental data, test the quality of chiral nuclear forces. We discuss applications of NCSMC to the $\alpha{-}\alpha$ scattering and the structure of $^8$Be, the p+$^7$Be and p+$^7$Li radiative capture and the production of the hypothetical X17 boson claimed in ATOMKI experiments. The $^7$Be(p,$\gamma$)$^8$B reaction plays a role in Solar nucleosynthesis and Solar neutrino physics and has been subject of numerous experimental investigations. We also highlight our investigation of the neutron rich exotic $^8$He that has been recently studied experimentally at TRIUMF with an unexpected deformation reported.
\end{abstract}

\section{Introduction}

The no-core shell model with continuum (NCSMC), first introduced in Refs.~\cite{Baroni2013L,Baroni2013}, is a  first-principles technique that has been successful in delivering predictive calculations of nuclear properties of light nuclei by combining bound and dynamic descriptions of an $A$-nucleon system (see Ref.~\cite{Navratil2016} for a review). In the NCSMC, the $A$-body Schr{\"o}dinger equation is solved for both bound and scattering boundary conditions with a trial wave function consisting of two parts: one describing the aggregate system when all nucleons are close together and the other considering explicitly sub-clusters that can completely separate. The input for these calculations are chiral Effective Field Theory (EFT) nucleon-nucleon (NN) and three-nucleon (3N) interactions.

\section{Many-body calculation of $\alpha{-}\alpha$ scattering}

\begin{figure}[h]
\begin{minipage}{18pc}
\includegraphics[width=18pc]{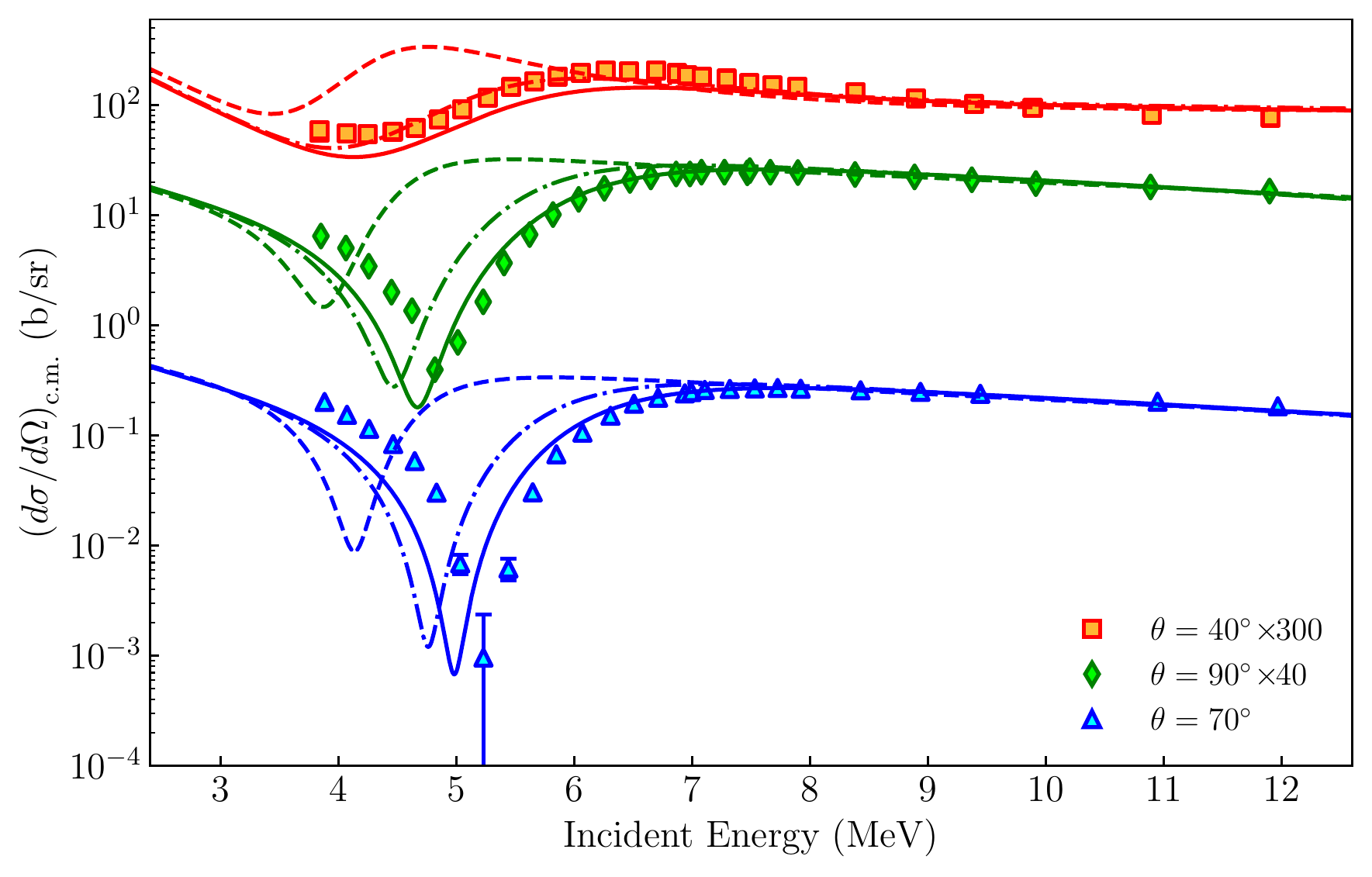}
\caption{\label{fig:alpha_alpha} The $\alpha{-}\alpha$ scattering 
differential cross section at c.m. angle $\theta{=}40^o, 90^o, 70^o$. 
The NN+3N$_{\rm lnl}$ Hamiltonian yields the overall best agreement with the data (symbols) from (a) Ref.~\cite{RevModPhys.41.247} and (b) Ref.~\cite{PhysRev.129.2252}. Further details are given in Ref.~\cite{Kravvaris:2020cvn}.}
\end{minipage}\hspace{1pc}%
\begin{minipage}{18pc}
\includegraphics[width=19pc]{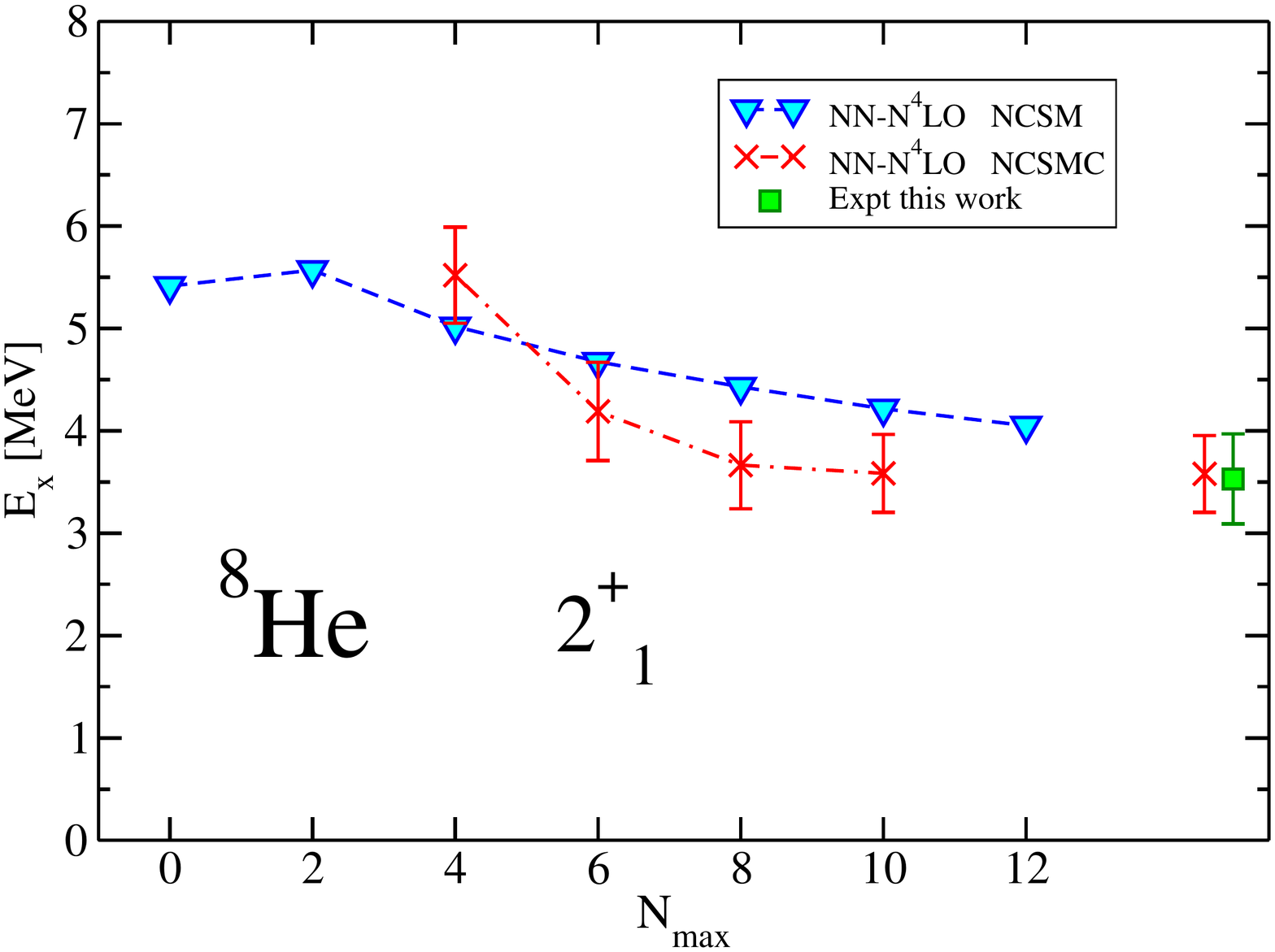}
\caption{\label{fig_NCSM_2p}$^8$He $2^+_1$ excitation energy dependence on the NCSM and NCSMC basis size. 
Extrapolated values and the experimental result~\cite{HOLL2021136710} are shown on the right. The vertical bars represent resonance widths. 
Further details are given in Ref.~\cite{HOLL2021136710}.}
\end{minipage} 
\end{figure}

A major milestone in our \emph{ab initio} reaction theory has been achieved recently by the development of a new formalism that takes full advantage of powerful second-quantization techniques, enabling the description of $\alpha{-}\alpha$ scattering and an exploration of clustering in exotic nuclei such as $^{12}$Be~\cite{Kravvaris:2020cvn}. Unlike in the original NCSMC formulation~\cite{Baroni2013,Navratil2016}, the cluster-cluster part now treats both the target and the composite projectile on the same footing utilizing the second quantization for constructing the reaction channels and evaluating their full matrix elements. The calculations are facilitated by the technique of the boost of the center-of-mass quanta described in Refs.~\cite{PhysRevLett.119.062501,PhysRevC.100.034321}. In Fig.~\ref{fig:alpha_alpha}, we compare the $\alpha{-}\alpha$ 
differential cross sections obtained with the NN interaction only and with NN+3N interactions contrasting two different 3N models. The best agreement with experimental data in particular in the region impacted by the $2^+_1$ resonance in $^8$Be is achieved with the NN+3N$_{\rm lnl}$ model~\cite{Entem2003,Soma2020}.

\section{The $2^+$ resonance of the exotic $^8$He}

The $^8$He is an exotic nucleus with an extreme neutron to proton ratio $N/Z{=}3$. It is the drip line nucleus of the helium isotopic chain and its stronger binding suggests a possible closed sub-shell at $N{=}6$ which would make $^8$He a doubly closed shell nucleus. Properties of this exotic system were investigate at TRIUMF ISAC facility. Proton inelastic scattering of $^8$He in inverse kinametics have been performed at the IRIS station~\cite{HOLL2021136710}. A resonance at 3.54(6) MeV with a width of 0.89(11) MeV was found. A coupled-channel and DWBA analysis identifies the resonance as a $2^+$ state in $^8$He.

\emph{Ab initio} NCSM and NCSMC calculations describe successfully the observed resonance~\cite{HOLL2021136710}. We employed the same Hamiltonian as in our recent investigation of $^9$He~\cite{PhysRevC.97.034314}. The NN interaction, denoted here as NN-N$^4$LO~\cite{Entem2017}, was renormalized by the SRG approach~\cite{Bogner2007} with an evolution parameter $\lambda_{\rm SRG}{=}2.4$~fm$^{-1}$. The NCSM and NCSMC convergence of the $^8$He $2^+_1$ excitation energy on the basis size characterized by $N_{\rm max}$ is shown in Fig.~\ref{fig_NCSM_2p}. 
The NCSM calculations yield a large quadrupole neutron moment $Q_n = 6.15\;e$~fm$^2$ and a small proton quadrupole moment, $Q_p= 0.60 \; e$~fm$^2$ for the 2$^+_1$ state. For $^{12}$C we predict $Q_n{\approx}Q_p{\sim}6 \; e$~fm$^2$. Thus, the neutron deformation in $^8$He is similar to that in $^{12}$C and qualitatively consistent with the experimental observations.

As the $^7$He $3/2^-$ ground state resonance is experimentally rather narrow (150 keV~\cite{TILLEY20023}), it is reasonable to use the $^7$He(gs)+n cluster to perform NCSMC calculations that extend the $^8$He NCSM basis. 
The resonance position and width were determined by analyzing the $2^+$ eigenphase shifts. The $2^+$ resonance appears in the $^{5}P_2$ partial wave. In the eigenphase shift, there is also a significant admixture of the  $^{3}P_2$ partial wave. The calculated $2^+_1$ excitation energy of 3.58(6) MeV and the width to be 750(50) keV is in an excellent agreement with the present experimental measurement.

\section{Radiative proton capture on $^7$Be}

Occurring at the tail end of the proton-proton chain, the radiative capture of a proton by a $^7$Be nucleus to produce an $^8$B nucleus (or $^7$Be(p,$\gamma$)$^8$B reaction) is key in determining the solar neutrino flux measured in terrestrial observatories.  Given its importance, it has been measured multiple times over the years with various techniques. However, due to Coulomb repulsion between the proton and the $^7$Be nucleus, a direct measurement at the astrophysically relevant energies is still missing, and theory calculations are used to extrapolate from higher energy experimental data. As a result of this extrapolation process, the uncertainty in the currently recommended~\cite{RevModPhys.83.195} value of the zero-energy S-factor, $S_{17}(0)=20.8\pm0.7\mathrm{(expt)}\pm1.4\mathrm{(theory)}$ eV$\cdot$barn, is dominated by theoretical contributions.

\begin{figure}[h]
\begin{minipage}{18pc}
\includegraphics[width=19pc]{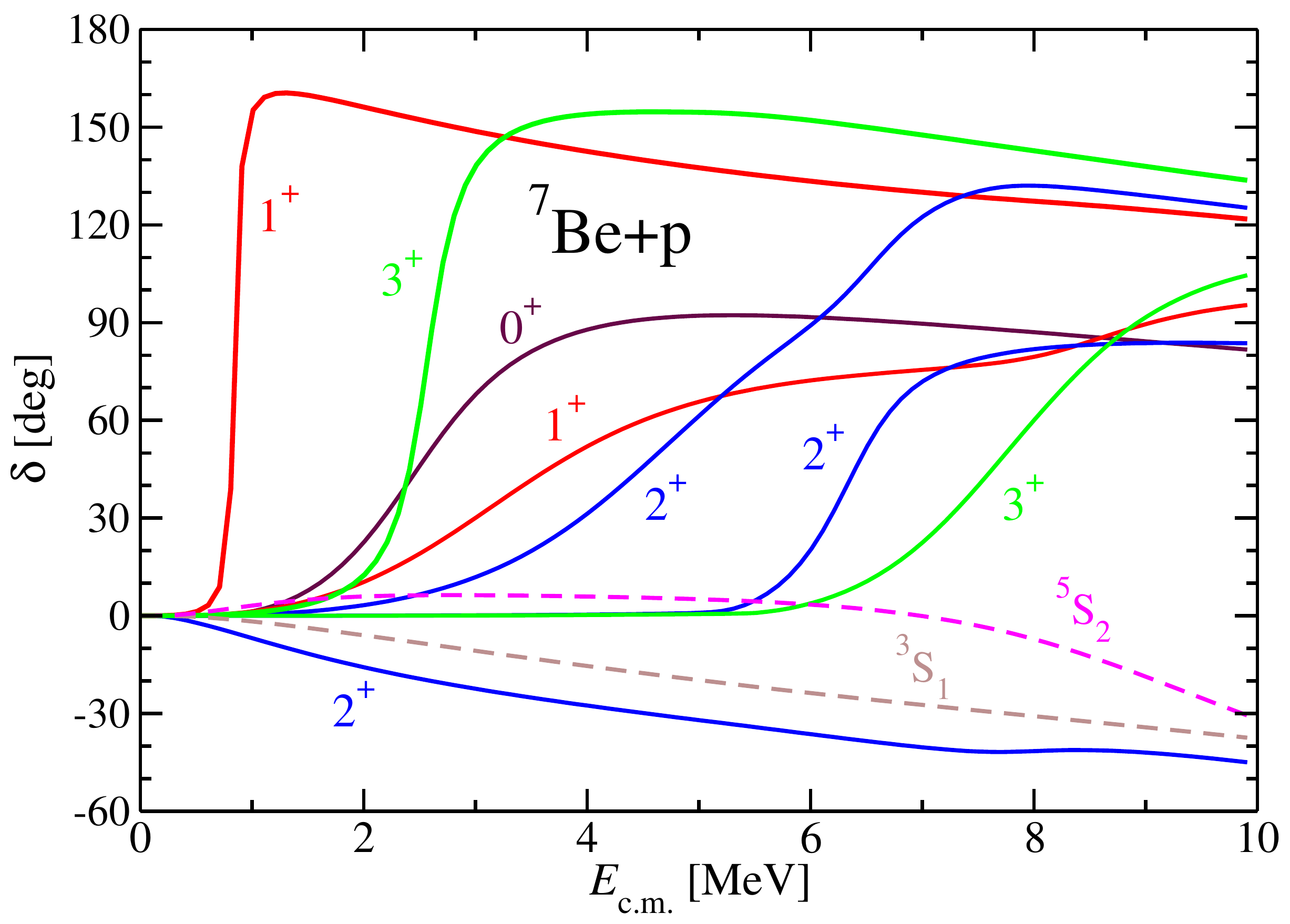}
\end{minipage}\hspace{1pc}%
\begin{minipage}{18pc}
\includegraphics[width=19pc]{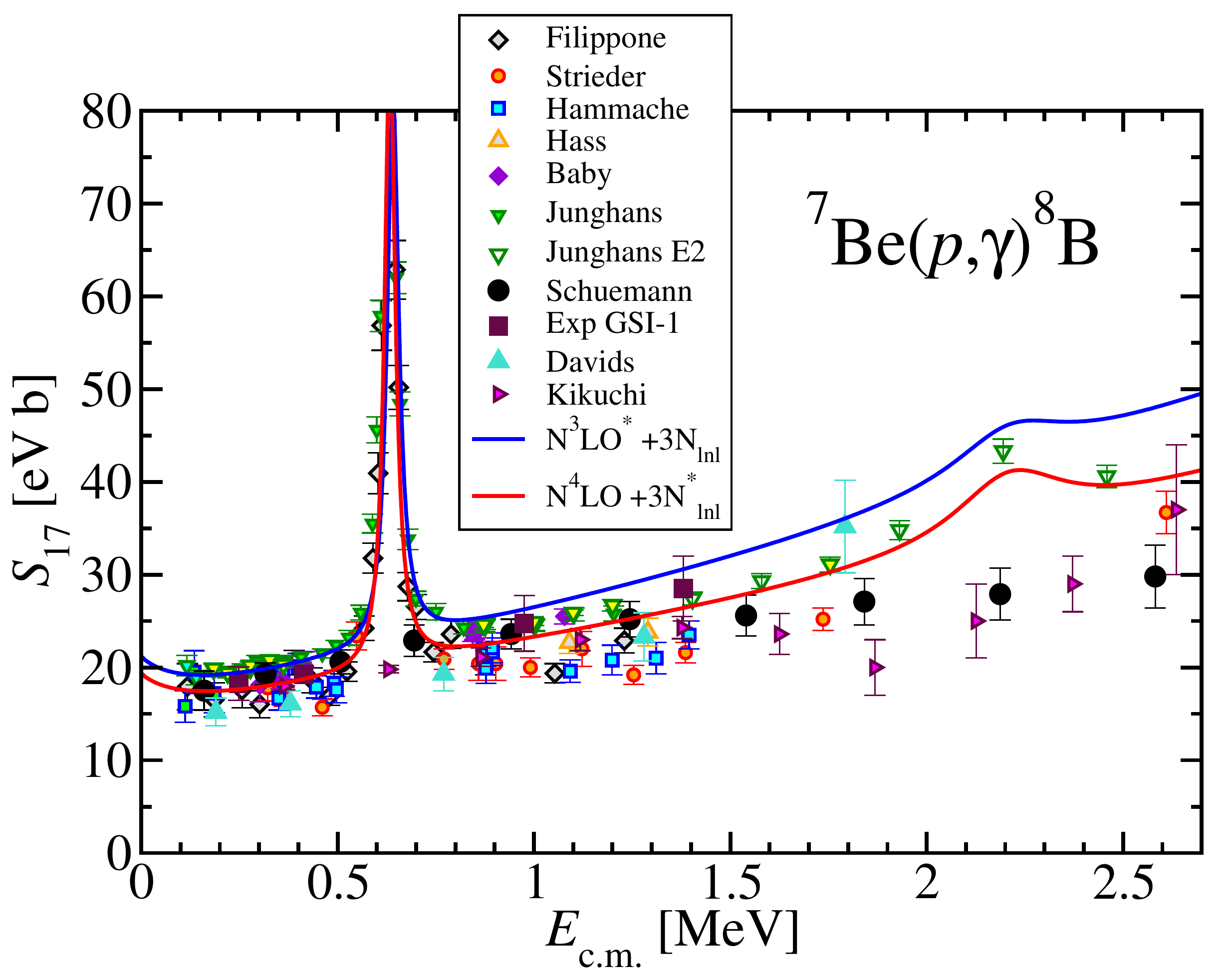}
\end{minipage}
\caption{\label{fig:Be7p_Sfact}Left panel: $^7$Be+p eigenphase shifts (solid lines) and $^3S_1$ and $^5S_2$ diagonal phase shifts (dashed lines) obtained from the NCSMC approach with the N$^4$LO+3N$^*_{\rm lnl}$ interaction. Right panel: Astrophysical S-factor of the $^7$Be(p,$\gamma$)$^8$B radiative capture obtained from the NCSMC approach with the N$^3$LO$^*$+3N$_{\rm lnl}$ (blue line) and the N$^4$LO+3N$^*_{\rm lnl}$ (red line) interactions compared to experimental data.  Details are given in Ref.~\cite{Kravvaris2022ab}.}
\end{figure}
In Ref.~\cite{Kravvaris2022ab}, we presented first-principle calculations of the $^7$Be+$p$ system, including the $^7$Be(p,$\gamma$)$^8$B reaction, using NN and 3N interactions derived from chiral EFT.  We used the NN interactions at 4th order of the chiral expansion defined in Ref.~\cite{Entem2003}, denoted N$^3$LO$^*$, and those at 3rd, 4th and 5th (N$^4$LO) order of Ref. \cite{Entem2017}.  These NN interactions were supplemented by the 3N interaction of Ref.~\cite{VanKolck94} with both local (3N$_\mathrm{loc}$)~\cite{Navratil2007,Gazit2019} and local plus non-local (3N$_\mathrm{lnl}$) regulators~\cite{Soma2020,PhysRevC.102.024616}. Finally, a 3N interaction 
with an added sub-leading contact term enhancing the strength of the spin-orbit interaction~\cite{Girlanda2011} was also employed (3N$_\mathrm{lnl}^*$). 

As an example of obtained results, the $^7$Be+p phase shifts are presented in Fig.~\ref{fig:Be7p_Sfact} together with the astrophysical S factor obtained using the chiral NN N$^3$LO$^*$+3N$_\mathrm{lnl}$ and NN N$^4$LO+3N$_\mathrm{lnl}^*$ interactions. The positive parity eigenphase shifts show the well-established $1^+_1$ and $3^+_1$ resonances as well as predictions of several other broader resonances. The calculated astrophysical S-factor reproduces well the resonance contributions due to the M1 and to a smaller extent to the E2 transitions from the $1^+$ resonance (the sharp peak at $\sim0.6$~MeV) and $3^+$ resonance (the bump at $\sim2.2$~MeV) to the weakly bound $2^+$ ground state of $^8$B. The calculated S-factor is in a reasonble agreement with the Junghans direct measurement data~\cite{PhysRevC.68.065803}. Overall, the new NCSMC calculations~\cite{Kravvaris2022ab} suggest value for the $^7$Be$(p,\gamma)^8$B S-factor at zero energy of 19.8$\pm$0.3 eV$\cdot$b which is consistent with the latest recommended value~\cite{RevModPhys.83.195} with the theoretical uncertainty significantly reduced. 

\section{Proton capture on $^7$Li and the X17 boson}

Anomalies in $^8$Be and $^4$He decay, in particular in the electron-positron pair production from the proton capture on $^7$Li and $^3$H were reported by the ATOMKI collaboration and interpreted by the decay of a new boson called X17 with the mass ${\sim}17$ MeV~\cite{PhysRevLett.116.042501,Firak2020, PhysRevC.104.044003,https://doi.org/10.48550/arxiv.2205.07744}.

\begin{figure}[h]
\begin{minipage}{18pc}
\includegraphics[width=18pc]{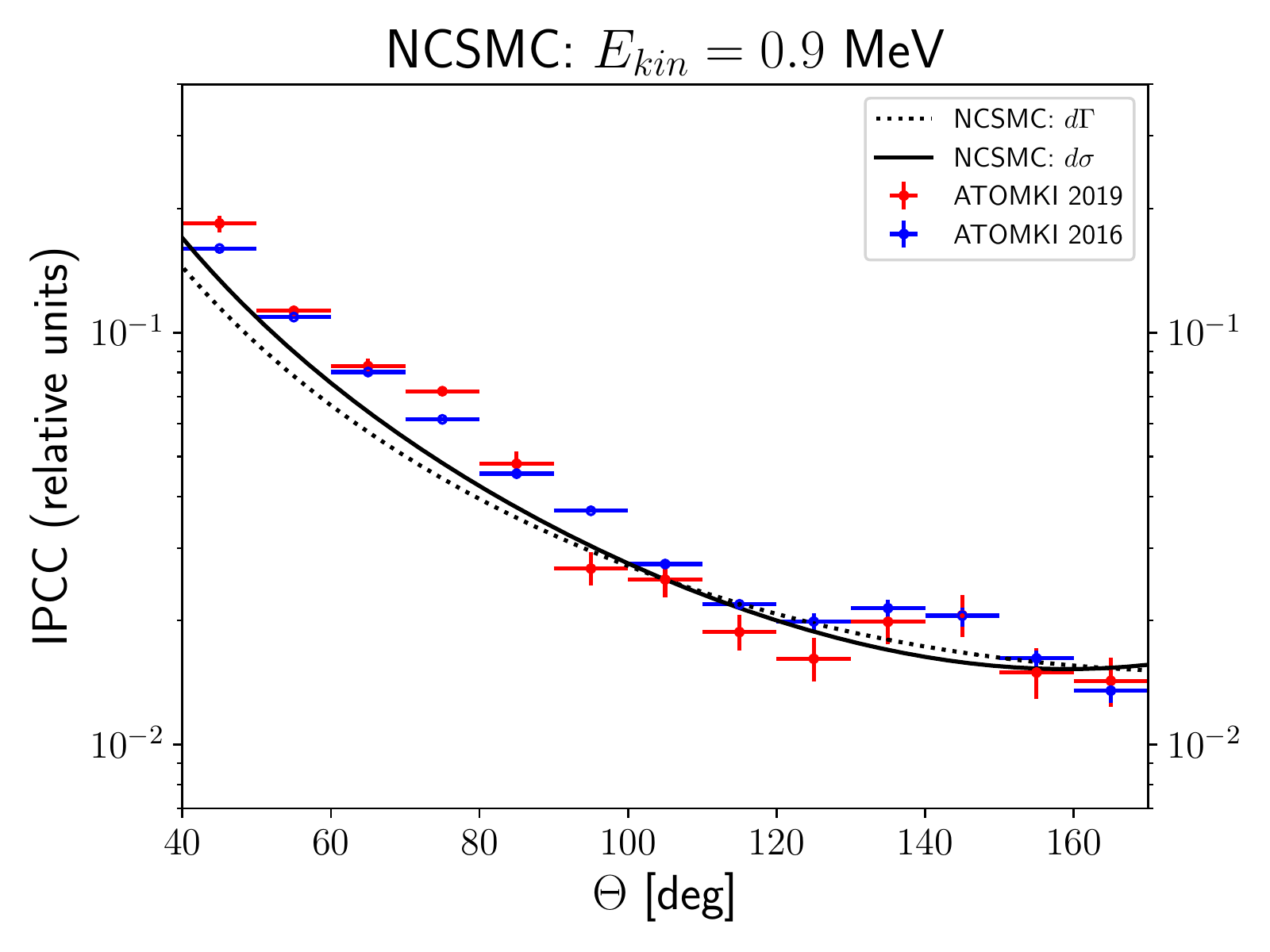}
\end{minipage}
\begin{minipage}{19pc}
 \includegraphics[width=20pc]{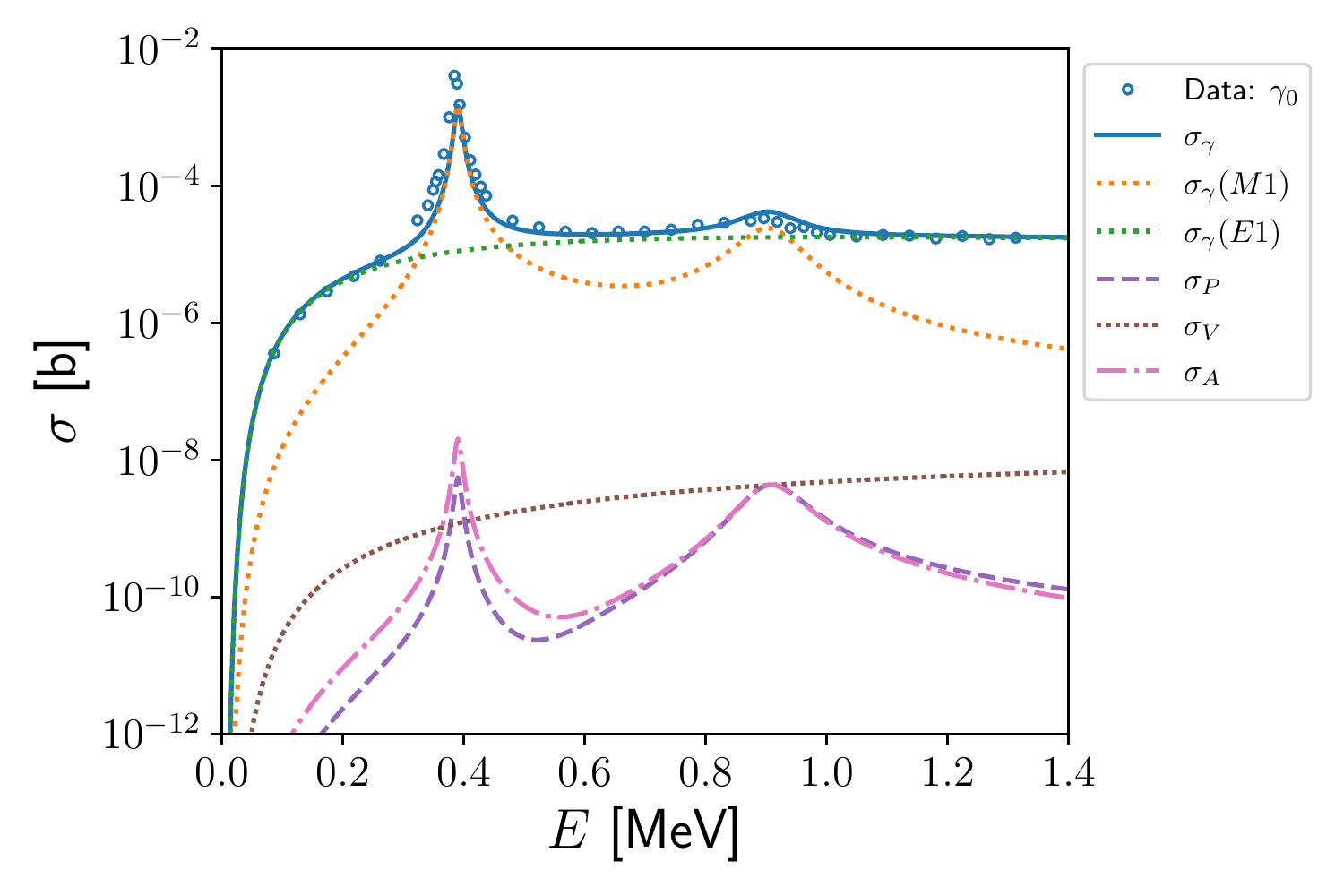}
 \end{minipage} 
\caption{\label{fig:Li7p_xsect}Left panel: Calculated internal pair conversion dependence on the angle between $e^+$ and $e^-$ matched to ATOMKI data at $105^o$ The dotted line shows results summing up the $E1$, $M1$, and $E2$ electromagnetic contributions. The solid line includes interference between the partial waves. Right panel: The proton capture cross section on $^7$Li dependence on the center-of-mass energy. The calculated $\gamma$ emission cross section is compared to experimental data from Ref.~\cite{Zahnow1995} and to the calculated hypothetical X17 boson emissions considering the $E1$ vector, pseudoscalar (axion), and axial vector boson candidates. All transitions are to the $^8$Be $0^+$ ground state. Calculations are preliminary, details will be given in Ref.~\cite{Gysbers2022}.}
\end{figure}
We embarked on an in-depth approach to the $p{+}^7$Li capture reaction within the {\it ab initio} NCSMC approach that allows us to describe simultaneously the structure of $^8$Be, the elastic and inelastic $^7$Li($p,p$) scattering, the charge exchange reaction $^7$Li($p,n$)$^7$Be, the $\gamma$ capture $^7$Li($p,\gamma$)$^8$Be, the internal pair conversion $^7$Li($p,e^+ e^-$)$^8$Be as well as the X17 boson production $^7$Li($p,X$)$^8$Be and decay for a variety of candidates for the hypothetical boson~\cite{Gysbers2022}. Preliminary NCSMC $^7$Li($p,e^+ e^-$)$^8$Be internal pair conversion correlation and integrated cross section results are shown in Fig.~\ref{fig:Li7p_xsect}. Summing up the $E1$, $M1$, and $E2$ electromagnetic (EM) contributions using the formalism of Ref.~\cite{PhysRevC.105.055502} we obtain the dotted line shown in left panel. Including interference of the partial waves following the formalism of Ref.~\cite{PhysRevC.105.014001} we obtain a better agreement with the ATOMKI EM background data. The calculated integrated $^7$Li($p,\gamma$)$^8$Be radiative capture cross section compares well with the data from Ref.~\cite{Zahnow1995}. The $E1$ and $M1$ contributions are shown separately. The two peaks dominated by the $M1$ contributions are due to the two $1^+$ resonances in $^8$Be with the lower one predominantly isospin $T{=}1$ while the higher one predominantly $T{=}0$ with a suppressed $M1$ decay to the $^8$Be $0^+$ $T{=}0$ ground state. Cross sections for the emission of the X17 boson lower by several orders of magnitude are also shown. Three X17 candidates were considered, a pseudo scalar (axion), axial vector, and $E1$ vector using operators from Refs.~\cite{PhysRevD.95.115024,PhysRevD.18.1607,ZHANG2021136061,PhysRevLett.128.091802}. The electron-positron pair production cross section (not shown in the figure) has a shape basically identical to the $\gamma$ capture cross section with its magnitude scaled by a factor of $\sim \alpha/2\pi \sim 10^{-3}$. One could understand then that an anomaly would be hardly observed in the first resonance with the very high electromagnetic $M1$ rate. An effect from the hypothetical boson can be expected in the second $1^+$ resonance where both the pseudoscalar and the axial vector boson candidate cross sections peak. An anomaly between the $1^+$ resonances would be consistent with the $E1$ vector. The latter is the preferred candidate according to the latest ATOMKI publications~\cite{PhysRevC.104.044003,https://doi.org/10.48550/arxiv.2205.07744}.

\section{Conclusions}

We presented several recent results in $A{=}8$ nuclei that demonstrate capabilities of the \emph{ab initio} NCSMC. With high-precision chiral NN+3N interactions as the input, one is able to predict with confidence properties of light nuclei even with a large neutron or proton excess. The method is capable to address issues of interest such as the evaluation of cross sections of reactions important for astrophysics or the X17 anomaly.

\ack
Prepared in part by LLNL under Contract DE-AC52-07NA27344.cSupport by the U.S.\ Department of Energy, Office of Science, Office of Nuclear Physics, under Work Proposal No.\ SCW0498, LLNL LDRD Project No. 20-LW-046, by the FRIB Theory Alliance award no. DE-SC0013617, and by the NSERC Grants SAPIN-2022-00019 and SAPPJ-2019-00039 is acknowledged.TRIUMF receives federal funding via a contribution agreement with the National Research Council of Canada. Computing support came from an INCITE Award on the Summit supercomputer of the OLCF at ORNL, from  the LLNL institutional Computing Grand Challenge program, and the Digital Research Alliance of Canada.

\section*{References}


\providecommand{\newblock}{}


\end{document}